# Response of the CALICE DHCAL to Hadrons and Positrons

Lei Xia[*]
On behalf of the CALICE collaboration

High Energy Physics Division, Argonne National Laboratory
9700 S Cass Ave, Argonne, IL 60439, USA

The Digital Hadron Calorimeter (DHCAL) $1m^3$ physics prototype was constructed and has been tested in particle beams. We report the preliminary results from the analysis of both positon and pion events of momenta between 2 and 60 GeV/c. These results are considered as a first validation of the DHCAL concept.

## 1 Introduction

After the successful construction and testing of the small size CALICE Digital Hadron Calorimeter (DHCAL) prototype, in the following named Vertical Slice Test (VST) [1–5], with 20 x 20 $cm^2$ Resistive Plate Chambers (RPCs), a larger prototype with 52 layers of 96 x 96 cm2 active area (1 x 1 $cm^2$ pads) was built in 2008-2010 [6]. 38 layers were the actual DHCAL and 14 layers were inserted into a tail catcher. With this large size technical prototype, we observe hadron showers with unprecedented spatial resolution. Among the various physics and technical measurements that the DHCAL offers, a first look at the data to investigate the hadronic and electromagnetic energy measurements with preliminary approaches is presented. Additional details on the large size prototype are given in [6] and the muon data analysis is presented in [7].

## 2 Analysis strategy

The data read out from the DHCAL contain the hit position information, the time stamp of the individual hits and the time stamp from the trigger. The hits in each layer are then combined into clusters using a nearest-neighbor clustering algorithm. If two hits share a common edge, they are assigned to the same cluster. The analysis presented in this paper considers only the first 38 layers of the DHCAL. The smallest layer number is the most upstream one. The event selection requires not more than 1 cluster in Layer 1 with least three active layers and no hits in the two outermost pads in any layer select the events with sufficient activity in the DHCAL and the events that have transverse containment, respectively.

The beam is a momentum selected mixture of muons, pions and positrons. This analysis is based on the topological particle identification utilizing the high segmentation of the DHCAL at beam momenta of 2, 4, 8, 12, 20, 25 and 32 GeV/c. Beam "momenta" and "energy" will be used interchangeably to refer to these numbers throughout this text.

---

[*] Speaker, however, the analysis was done by Burak Bilki, Argonne/University of Iowa.

The Muon identification is based on finding a linear alignment of isolated clusters within the DHCAL. The cluster in Layer 1 is matched with the clusters in the last layers and a line joining these clusters is formed. Then for all the layers in between, clusters with a distance less than 2 cm from the line are searched for. These clusters are required to be isolated from 1.5 cm to 25 cm from the predicted point. If the total number of layers with aligned clusters is equal to the total number of active layers (layers with at least one hit), the track is identified as a muon. This case constitutes more than 60% of all the muons in a typical muon run. The remaining 40% of the muons have a few higher multiplicity layers along the track which violate the isolation requirement. These high multiplicities are due to delta ray production and are not correlated between the layers. In order to tag these muon tracks, events with total number of layers with aligned clusters exceeding 80% of the number of active layers are considered. If the isolation criterion is not violated in two consecutive layers, the track is assigned a muon ID. Figure 1 demonstrates the application of the muon identification algorithm to one of the muon runs. The algorithm is 97% efficient in identifying muons. The disagreement in the distributions is mainly in the higher end of the spectrum indicating that these muons have delta ray production in at least two consecutive layers or the tracks are highly inclined, hence the higher number of hits. 95% of the remaining, unidentified muons are excluded from the pion and positron analyses by requiring no hits in the last two layers of the DHCAL (longitudinal confinement requirement).

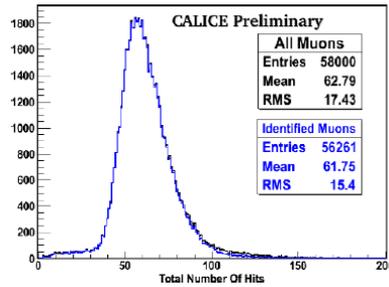

Fig. 1. Distribution of the number of hits in muon data selected with the Muon-ID algorithm (blue histogram). The inclusive muon data is shown in black.

Once the muons are identified, the particle identification proceeds with finding the pions. First, the MIP segment is identified by a similar algorithm to the muon identification. Once the last MIP layer is determined (with having less than 4 hits), track segments from this layer towards the back of the DHCAL are searched for (track segment is defined as the aligned clusters in at least three layers). These track segments should not be aligned with the MIP segment. If any track segment that extends to at least four layers is found, the particle is identified as pion. For the track segments that extend only to three layers, the angle between these track segments is considered. If a pair with more than 20° angle in between is found, the particle is assigned a pion ID. These selections are based on the event topology in the interaction region of the pions. At the end of this step, the unidentified events contain all the positrons and a fraction of the pions. A transverse size parameter is defined to distinguish between the pions and the positrons in the remaining sample: $r_{rms} = \sqrt{\frac{\sum r_i^2}{N_{hits}}}$, where $r_i$ is the distance of each hit to the x-y center of all the hits in the corresponding layer and $N_{hits}$ is the total number of hits. Figure 2 shows the distribution of the $r_{rms}$ variable for the 20 GeV/c runs. A selection cut at $r_{rms}=5$ discriminates between the pions ($r_{rms} >5$) and positrons ($r_{rms} <5$) based on the differences between the transverse sizes of electromagnetic and hadronic showers.

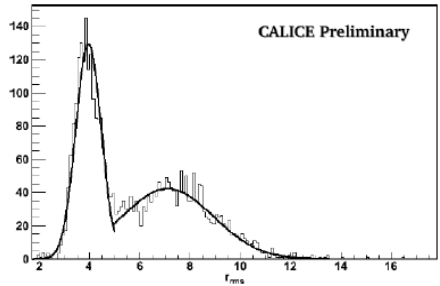

Fig. 2. $r_{rms}$ variable for the 20 GeV/c runs. Pions (positrons) tend to populate high (low) values of $r_{rms}$.

Figures 3 a-n show the distribution of the total number of hits for pions and positrons after particle identification and longitudinal confinement requirement.

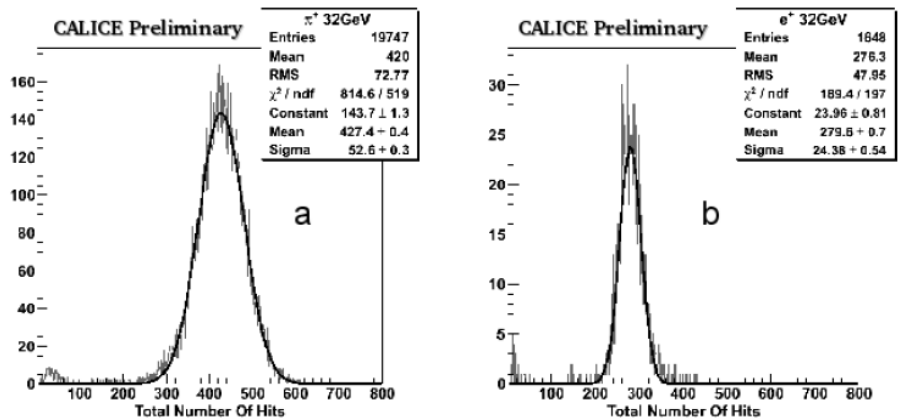

Fig. 3. Distribution of total number of hits for 32, 25, 20, 12, 8, 4 and 2 GeV pions (a, c, e, g, i, k, m) and positrons (b, d, f, h, j, l, n) and the Gaussian fits.

At higher beam energies, the fraction of positrons in the beam is lower when compared to lower energies and the total hits spectra are well separated. At lower beam energies, the spectra overlap and the event topology has less discriminative power. Therefore the distributions in Figs. 3 k, m are still significantly contaminated by positrons and unidentified low energy muons that have a higher probability of multiple scattering when compared to high energy ones. As additional tools and more sophisticated algorithms are being developed, 4 GeV and 2 GeV pion responses are not included in the current measurements.

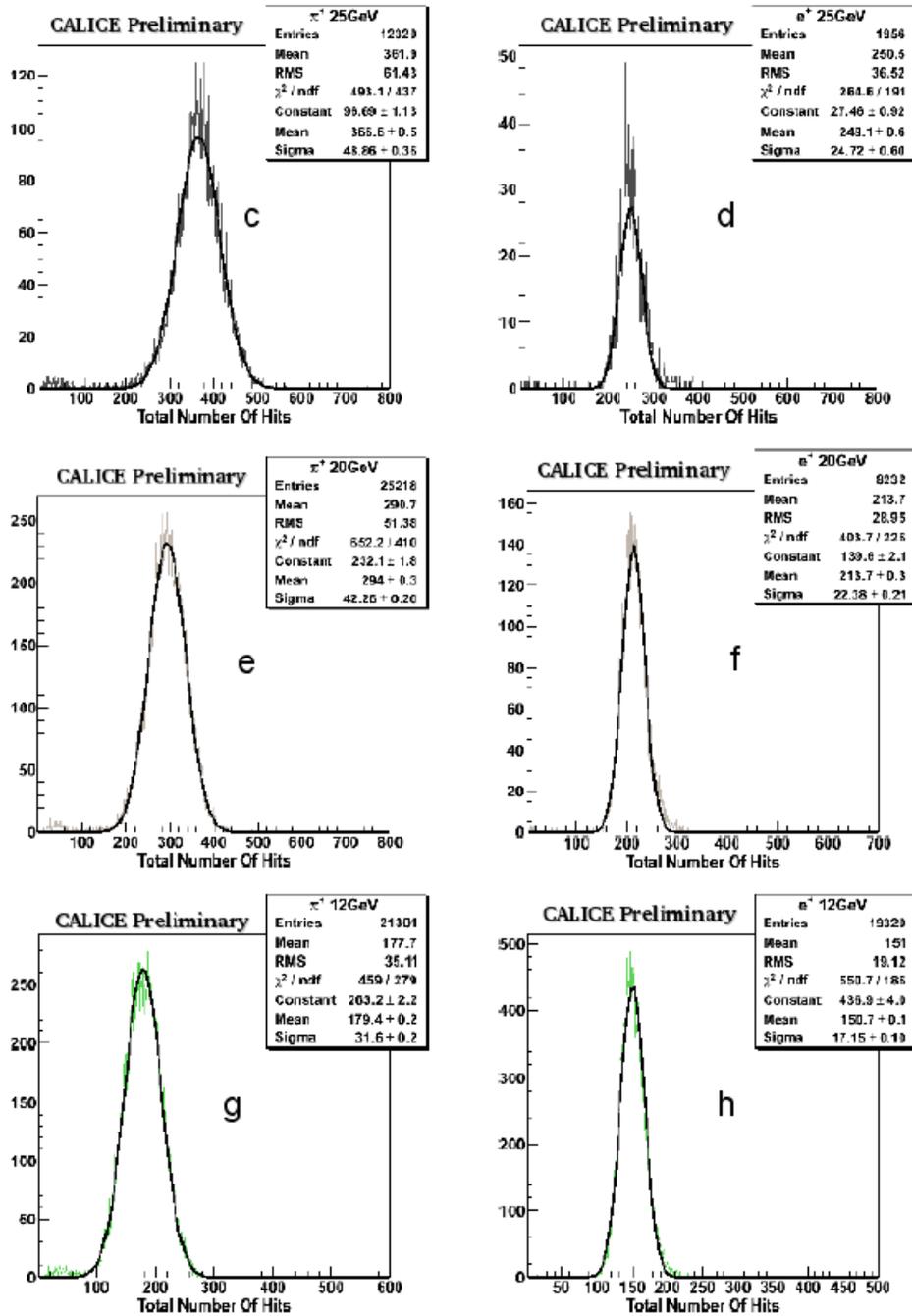

Fig. 3 Distribution of total number of hits for 32, 25, 20, 12, 8, 4 and 2 GeV pions (a, c, e, g, i, k, m) and positrons (b, d, f, h, j, l, n) and the Gaussian fits. - continued.

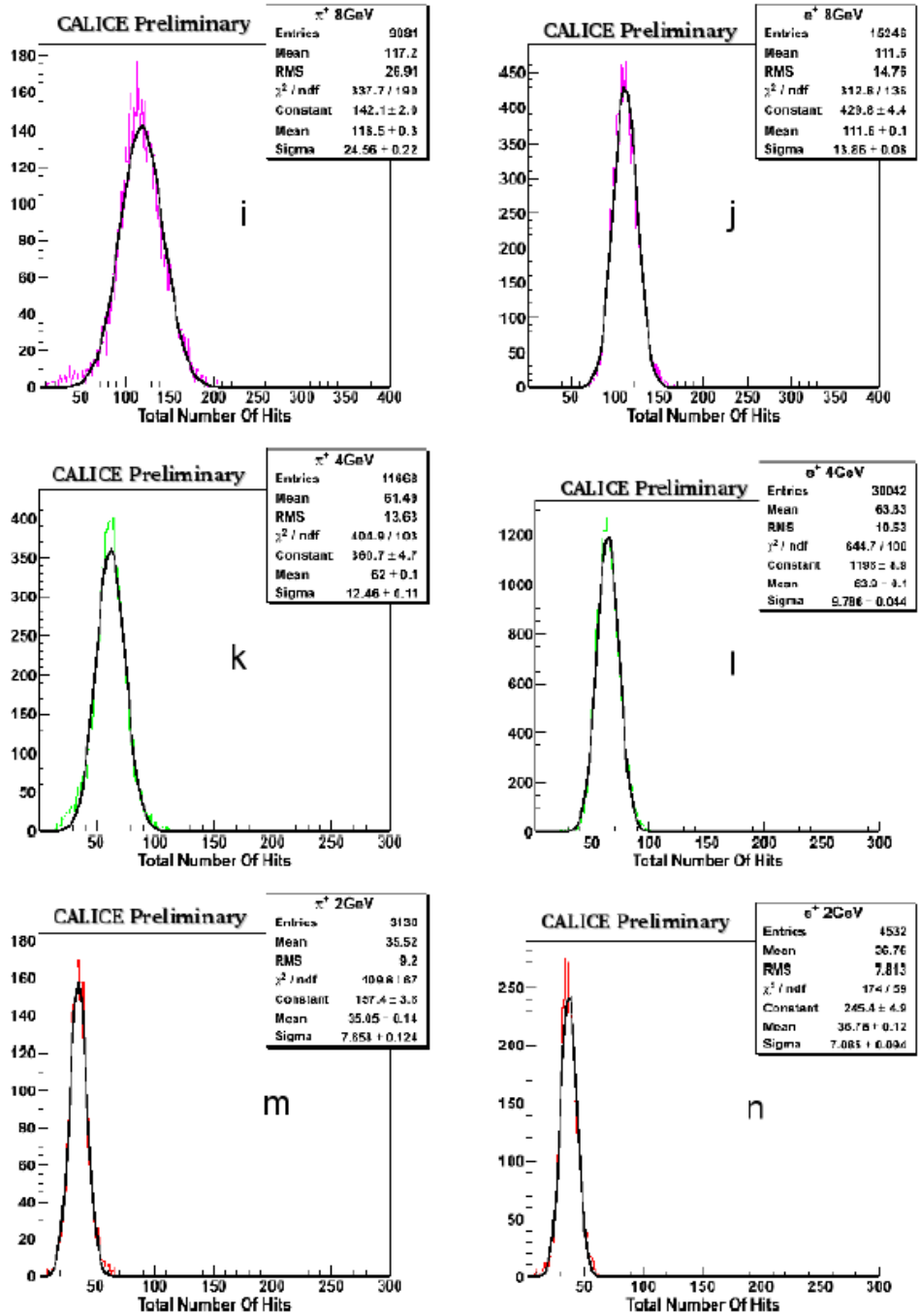

Fig. 3 Distribution of total number of hits for 32, 25, 20, 12, 8, 4 and 2 GeV pions (a, c, e, g, i, k, m) and positrons (b, d, f, h, j, l, n) and the Gaussian fits. - continued.

## 3  Response of DHCAL to Hadrons

The mean response of the DHCAL to pions is shown in Fig. 4. The response is linear up to 25 GeV, and at 32 GeV, the response deviates from linear behavior due to RPC response fluctuation and possible saturation effect. Therefore, 32 GeV data point is not included in the linear fit (N=aE where N is the total number of hits and E is the beam energy). Figure 4 also shows the percent difference between measured and fit values. Except the 32 GeV points, the deviation is within 2%. Note that these results have been obtained without a calibration of the response of the DHCAL.

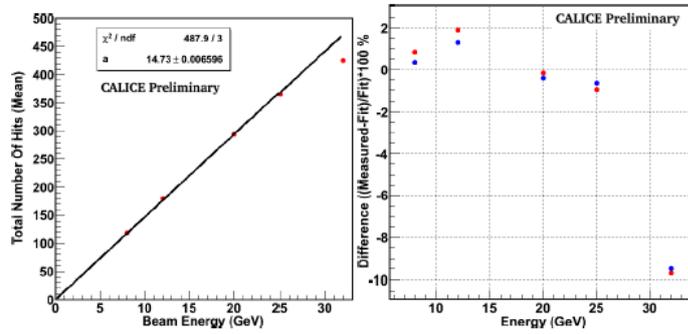

Fig. 4. The mean response of DHCAL to pions of 8, 12, 20, 25 and 32 GeV energy (left) and the difference between the measured and fit values for all identified pions (red) and longitudinally contained pions (no hits in the last two layers, blue) (right).

Figure 5 shows the hadronic energy resolution of the DHCAL with the current particle identification algorithms. 32 GeV points are excluded from the fits to the function $\frac{\sigma}{E} = \frac{\alpha}{\sqrt{E}} \oplus C$ where α is the stochastic term and C is the constant term. The fits represent the data well and for the longitudinally contained pions -that have no hits in the last two layers- a stochastic term of approximately 55% and a constant term of 7.5% is achieved. The measurements are within 1-2% of predictions based on the simulation of the large-size DHCAL prototype using the VST results [2].

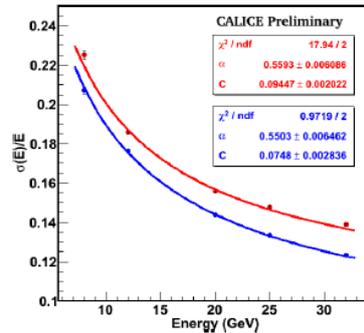

Fig. 5. Hadronic energy resolution of DHCAL for all identified pions (red) and the longitudinally contained pions (blue).

## 4 Response of DHCAL to Positrons

The mean response of the DHCAL to identified positrons is shown in Fig. 6. The response is fit with the nonlinear function $N=a+bE^m$. The fit describes the data well and is in accordance with the predictions in the VST results of positron showers [4]. In order to measure the electromagnetic energy resolution of the DHCAL with the positron response corrected for non-linearity, the total number of hits for each positron event is mapped into its corresponding energy value using the fit function in Fig. 6. Then these reconstructed energy spectra are used to calculate the energy resolutions. Figure 7 shows the results of the energy reconstruction for 4 GeV (a) and 12 GeV (b) positron runs together with the Gaussian fits. Figure 8 shows the electromagnetic energy resolution for both uncorrected and corrected values.

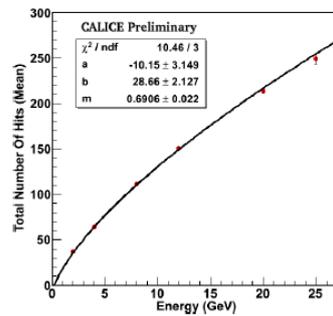

Fig. 6. Mean response of DHCAL to positrons at 2, 4, 8, 12, 20 and 25 GeV.

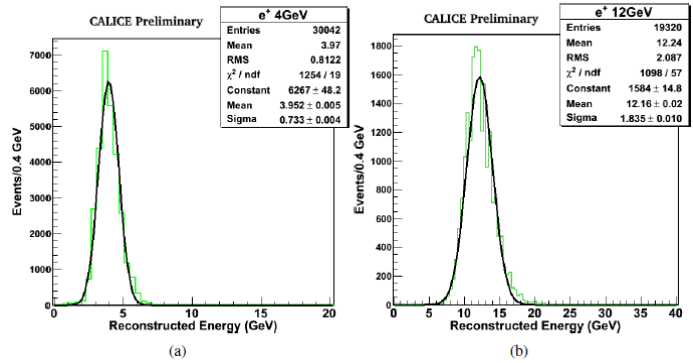

Fig. 7. Reconstructed energy spectra for 4 GeV (a) and 12 GeV (b) positrons.

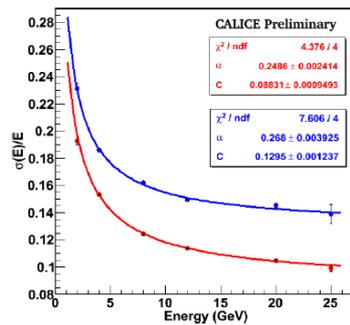

Fig. 8. Corrected (blue) and uncorrected (red) electromagnetic energy resolution for DHCAL at 2, 4, 8, 12, 20 and 25 GeV.

LCWS11

## 5  Summary

With the several successful test beam campaigns since October 2010, the digital hadron calorimeter concept is being validated under extensive physics and technical tests. This paper presents a first look analysis on the October 2010 data to obtain the digital hadron calorimeter properties. More sophisticated analyses will be forthcoming in the near future.

The particle identification algorithms defined within this text provide sufficiently well discrimination at high energies. However, the complications in the event topologies at low energies require further studies to integrate these energies into the calorimetric measurements. These new algorithms are expected to improve the current measurements as well. With the present algorithms, a hadronic energy resolution of $\frac{\sigma}{E} = \frac{55\%}{\sqrt{E}} \oplus 7.5\%$ and an electromagnetic energy resolution between 24% and 14% in the energy range of 2 to 25 GeV are obtained.

Further methods are being developed to obtain unbiased samples of pure beam particles and to obtain the DHCAL response not only as an energy measuring calorimeter, but also as a unique source of information of detailed hardonic interactions with unprecedented spatial resolution.

## 6  References

[1] Q. Zhang et.al., Environmental Dependence of The Performance of Resistive Plate Chambers, JINST 5 P02007, 2010.
[2] B. Bilki et.al., Hadron Showers in a Digital Hadron Calorimeter, JINST 4 P10008, 2009.
[3] B. Bilki et.al., Measurement of The Rate Capability of Resistive Plate Chambers, JINST 4 P06003, 2009.
[4] B. Bilki et.al., Measurement of Positron Showers With a Digital Hadron Calorimeter, JINST 4 P04006, 2009.
[5] B. Bilki et.al., Calibration of a Digital Hadron Calorimeter With Muons, JINST 3 P05001, 2008.
[6] K. Francis, Construction of The DHCAL, these proceedings.
[7] J. Repond, Analysis of Muon Events in The DHCAL, these proceedings.